\documentclass[aps,prl,twocolumn,showpacs,nosuperscriptaddress,groupedaddress]{revtex4}  % for review and submission
\usepackage{graphicx}  % needed for figures
\usepackage{bm}        % for math
\usepackage{amssymb}   % for math
\usepackage{amsmath}   % for line breaking
\usepackage{epstopdf}
\begin{document}

\title{Public good diffusion limits microbial mutualism}
\author{Rajita Menon}
\affiliation{Department of Physics, Boston University, Boston, Massachusetts 02215, USA}
\author{Kirill S. Korolev}
\affiliation{Department of Physics and Graduate Program in Bioinformatics, Boston University, Boston, Massachusetts 02215, USA}
\email{korolev@bu.edu}
\date{\today}

\begin{abstract}
Standard game theory cannot describe microbial interactions mediated by diffusible molecules. Nevertheless, we show that one can still model microbial dynamics using game theory with parameters renormalized by diffusion. Contrary to expectations, greater sharing of metabolites reduces the strength of cooperation and leads to species extinction via a nonequilibrium phase transition. We report analytic results for the critical diffusivity and the length scale of species intermixing. Species producing slower public good is favored by selection when fitness saturates with nutrient concentration. 
\end{abstract}

\maketitle

Complex microbial communities are essential for the environment and human health. Microbial functions range from the production of biofuels and the release of powerful greenhouse gasses to the production of cheese and the digestion of food inside our guts. Most of these functions are orchestrated by complex microbial consortia rather than single species~\cite{mee2014, oficteru2010}. To create and control such multispecies ecosystems, we need to understand the mechanisms that govern microbial coexistence and cooperation.

Heterotrophic cooperation is a common and perhaps the simplest element of complex microbial communities~\cite{mee2014, muller2014, momeni2013, momeni2013spatial}. In this two-way cross feeding, each species produces an amino acid or other metabolite necessary for the other species. Heterotrophic cooperation has been previously described by the evolutionary game theory~\cite{korolev2011, dallasta2013}, which assumes that microbes interact only with their closest neighbors. However, unlike human societies or bee colonies, microbial communities rarely rely on direct contact. Instead, microbes primarily communicate though diffusible molecules, which rapidly spread in the environment~\cite{west2007, kummerli2014, cordero2012, nadell2010, julou2013}. Because of this diffusive sharing within or between species, such molecules are often termed public goods. Broad understanding of how public good diffusion affects heterotrophic cooperation is still lacking. 

In this Letter, we explicitly account for the production, consumption, and diffusion of public goods in a model of heterotrophic cooperation. We find that unequal diffusivities of the public goods can significantly favor one of the species and even destroy their cooperation. More importantly, the diffusion of public goods has the opposite effect compared to species migration. Higher migration improves mutualism and stabilizes species coexistence. In contrast, cooperation is lost above a critical diffusivity of public goods, for which we obtain an analytical expression. We also describe the effect of public good diffusion on the spatial distribution of species, which is often used to quantify microbial experiments~\cite{muller2014, momeni2013, momeni2013spatial, korolev2011amnat}. Our analytical approach is based on computing how public good diffusion renormalizes the strength of selection and thus should be applicable to other models.
   
Motivated by the experiments on cross-feeding mutualists~\cite{muller2014, momeni2013, momeni2013spatial}, we consider two species (or strains)~A and~B producing public goods of type~$A$ and~$B$, respectively, and consuming public goods of the opposite type. These species live in a one-dimensional habitat, which corresponds to the quasi-one-dimensional edge of microbial colonies, where cells actively divide~\cite{korolev2010}. In simulations, the habitat is an array of islands populated by~$N$ cells each. This finite carrying capacity sets the magnitude of demographic fluctuations typically termed genetic drift~\cite{korolev2010}. Nearest-neighbor islands exchange migrants at a rate~$m$, which specifies the degree of movement within a microbial colony. In the continuum limit, the evolutionary dynamics of the species is described by

\begin{equation}
	\frac{\partial f_{\rm{A}}}{\partial t} = \frac{m}{2}\frac{\partial^{2} f_{\rm{A}}}{\partial x^{2}} + (w_{\rm{A}}-w_{\rm{B}}) f_{\rm{A}}f_{\rm{B}} + \sqrt{\frac{f_{\rm{A}}f_{\rm{B}}}{N}} \eta (t,x),
\label{eq:ssm}
\end{equation}

\noindent where~$t$ and~$x$ are time and position measured in such units that generation time and island spacing are set to~$1$; $f_{\rm{A}}(t,x)$ and~$f_{\rm{B}}(t,x)=1-f_{\rm{A}}(t,x)$ are the relative abundances of species~A and~B;~$w_{\rm{A}}$ and~$w_{\rm{B}}$ are the fitnesses of species~A and~B, respectively, that depend on the local concentration of the public goods; and~$\eta(t,x)$ is a delta-correlated Gaussian white noise. Equation~(\ref{eq:ssm}) represents the classical stepping-stone model of population genetics~\cite{korolev2010, kimura1964} and accurately describes population dynamics in microbial colonies~\cite{korolev2011amnat, frey_2010}. 

Standard game-theory treatments of microbes assume that the fitnesses~$w_{\rm{A}}$ and~$w_{\rm{B}}$ depend on the local abundances of the species rather than the public goods concentrations~\cite{korolev2011}. Here, we relax this assumption and consider

\begin{equation}
	\begin{aligned}
	&	w_{\rm{A}} = 1 + \frac{n_{{B}}}{1 + n_{{B}}/K_{B}}, \\
	&	w_{\rm{B}} = 1 + \frac{n_{{A}}}{1 + n_{{A}}/K_{A}}, \\
	\end{aligned}
\label{eq:fitness} 
\end{equation}

\noindent where~$n_{{A}}$ and~$n_{{B}}$ are the concentrations of the public goods measured in the units of fitness. 

In the simplest model, the dynamics of the public goods concentrations are given by the following reaction-diffusion equation:

\begin{equation}
\frac{\partial n_{{A}}}{\partial t} = D_{{A}}\frac{\partial^{2}n_{{A}}}{\partial x^{2}} + p_{{A}} f_{\rm{A}} - d_{{A}} n_{{A}},
\label{eq:pg} 
\end{equation}

\noindent and an analogous equation for~$n_{{B}}$. Here, for the public good of type~A, $D_{{A}}$ is the diffusivity,~$p_{{A}}$ is the production rate, and~$d_{{A}}$ is the rate of loss comprised of consumption by both species, spontaneous decay or degradation, and transport outside the region of microbial growth~\cite{muller2014}. Both~$p_{{A}}$ and~$d_{{A}}$ can depend on~$n_{{A}}$,~$n_{{B}}$ and~$f_{\rm{A}}$ in a more realistic model, but our simulations suggest that all important aspects of population dynamics are already captured by Eq.~(\ref{eq:pg})~\cite{supplement}. Since public good dynamics occurs much faster than cell migration and growth, public good concentrations equilibrate rapidly, i.e.~$\partial n_{{A}}/\partial t\approx\partial n_{{B}}/\partial t\approx0$. This results in

\begin{equation}
	n_{{A}} (x) = \frac{p_{{A}}}{2\sqrt{D_{{A}} d_{{A}}}} \int f_{\rm{A}}(x') e^{-\sqrt{\frac{d_{{A}}}{D_{{A}}}}|x' - x|}dx',
\label{eq:pg_sol} 
\end{equation}

\noindent and similarly for~$n_{{B}}$. Equations~(\ref{eq:ssm})--(\ref{eq:pg_sol}) have been previously used to simulate cooperatively growing microbial communities~\cite{muller2014, borenstein2013}, but analytical results and broad understanding of the effect of public good diffusion on population dynamics is still lacking. 

\begin{figure}[h]
\includegraphics[width=\columnwidth]{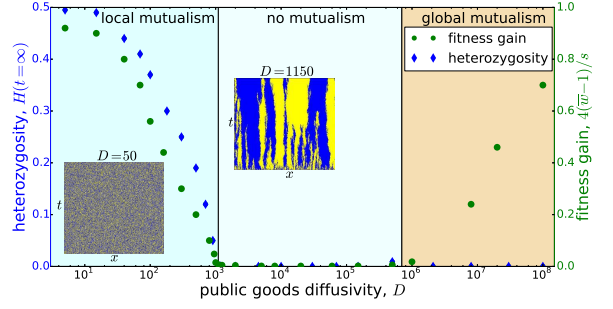}
\caption{Local heterozygosity decreases with the diffusivity of the public goods, and the species coexistence is completely lost above a critical diffusivity~$D_{\rm{c}}=1200$. Fitness gain due to mutualism also decreases with~$D$ and vanishes at~$D_{\rm{c}}$ until the diffusion becomes so large that nutrients can spread through the entire population and mutualism emerges on a global scale. The two insets show how the distribution of species labeled by different colors varies with position~(x-axis) and time~(y-axis) for small diffusivities~(left) and just below the critical diffusivity~(right). Note that the loss of mutualism observed here is triggered by changing~$D$ rather than the strength of selection or exogenous public goods concentrations as was done previously~\cite{korolev2011, muller2014}. \label{fig1}} 
\end{figure}

\begin{figure}[h]
\includegraphics[width=\columnwidth]{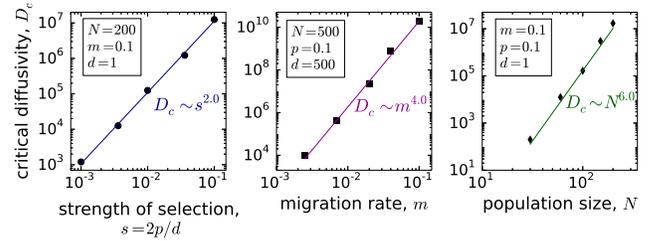}
\caption{Simulations~(points) confirm our analytical predictions~(lines) that critical diffusivity increases as~$s^{2}m^{4}N^{6}$. \label{fig2}} 
\end{figure}

To understand the overall effect of public good diffusion on microbial mutualism, it is sufficient to consider a simple symmetric case:~$p_{{A}}=p_{{B}}=p$, $d_{{A}}=d_{{B}}=d$, $D_{{A}}=D_{{B}}=D$, and $K_{{A}}=K_{{B}}=K=\infty$, which we proceed to analyze by combining Eqs.~(\ref{eq:ssm}), (\ref{eq:fitness}), and~(\ref{eq:pg_sol}):

\begin{equation}
	\begin{aligned}
		\frac{\partial f_{\rm{A}}}{\partial t} &=\frac{pf_{\rm{A}}(1-f_{\rm{A}})}{2\sqrt{Dd}} \int [1-2f_{\rm{A}}(x')] e^{-\sqrt{\frac{d}{D}}|x' - x|}dx' \\ &+\frac{m}{2}\frac{\partial^{2} f_{\rm{A}}}{\partial x^{2}} + \sqrt{\frac{f_{\rm{A}}(1-f_{\rm{A}})}{N}} \eta (t,x).
\label{eq:fmodel}
\end{aligned}
\end{equation}

For small~$D$, the integrand in the first term of Eq.~(\ref{eq:fmodel}) is peaked around~$x'=x$, so one can expand~$f_{\rm{A}}(x')$ in Taylor series around~$x$; see~\cite{supplement}. To the zeroth order, the selection term becomes~$sf_{\rm{A}}(1-f_{\rm{A}})(1/2-f_{\rm{A}})$, where~$s=2p/d$ is the strength of local frequency-dependent selection. Thus, in the limit~$D\to0$, our model of mutualism reduces to the standard game theory formulation with frequency-dependent selection~\cite{korolev2011, frey_2010}. Population dynamics in this limit are controlled by a dimensionless quantity~$S=smN^2$, which we refer to as the strength of the mutualism. When~$S$ exceeds~$S_{\rm{c}}$, a critical value of order~$1$, the mutualism is stable, and the two species coexist~\cite{korolev2011, supplement, durand1997}. In contrast, when~$S<S_{\rm{c}}$, selection for coexistence is not strong enough to overcome local species extinctions due to genetic drift, and the population becomes partitioned into domains exclusively occupied by one of the two species. With the loss of the coexistence, the mutualism also ceases because spatially segregated species cannot cooperate when public goods do not diffuse.

Since public good diffusion facilitates nutrient exchange between the segregated domains, one might naively expect that increasing~$D$ would rescue mutualism. We, however, find that higher nutrient diffusivities promote species demixing, and the separation between the domains grows with~$D$ much more rapidly than the distance over which public goods are exchanged. As a result, increasing the diffusion of public goods destroys mutualism.  

Because mutualism and coexistence are not equivalent when~$D>0$, we considered them separately. The benefit that species derive from the mutualism was quantified with the fitness gain~$4(\bar{w} - 1)/s$, where~$\bar{w}$ is the mean fitness of the population. When mutualism is lost, the mean fitness equals~$1$, and the fitness gain is~$0$, while successful mutualism results in~$\bar{w}\approx1+p/(2d)$ and the fitness gain of~$1$. Species intermixing was quantified by the average local heterozygosity~$H(t)=\langle 2f_{\rm{A}}(t,x)f_{\rm{B}}(t,x) \rangle$, which equals~$1/2$ for strongly intermixed species and~$0$ for species that are spatially segregated.

In simulations, we found that both fitness gain and local heterozygosity~$H$ decrease with~$D$, and mutualism is lost for diffusivities above a certain value~$D_{\rm{c}}$ (Fig.~\ref{fig1}). While~$H=0$ for all~$D>D_{\rm{c}}$, the fitness gain becomes nonzero for extremely large nutrient diffusivities, when the nutrient diffusion length scale becomes comparable to the system size. In this regime, nutrients concentrations are homogeneous across the entire population and species cooperate on a global rather than local scale. We discuss the transition to global mutualism further in the Supplemental Material~\cite{supplement} and show that global mutualism cannot occur in microbial populations because it requires nutrient diffusivities that are unrealistically high.

Note that the loss of mutualism in our model is due to genetic drift rather than the proliferation of nonproducers~(cheaters). In intraspecific cooperation, public good diffusion allows cheaters to consume nutrients that they do not produce and thus inhibits cooperation by a mechanism distinct from the one described here~\cite{allen_2013}.

To show how the interplay between population demixing and nutrient exchange leads to the loss of the mutualism, we introduce three length scales that control population dynamics. The first length scale appears in Eq.~(\ref{eq:pg_sol}) and relates the spatial variation in species abundances to the distribution of nutrients. This nutrient length scale,~$L_{\rm{n}}$, is the typical distance that nutrients diffuse before being consumed, and is given by $L_{\rm{n}}=\sqrt{D/d}$ since~$1/d$ is the typical nutrient lifetime. The second length scale describes species segregation. We define it as the size of the domains occupied by a single species~(insets in Fig.~\ref{fig1}) and denote it as~$L_{\rm{d}}$. Below, we show that~$L_{\rm{d}}$ increases with~$D$ much more rapidly than $L_{\rm{n}}$. The final length scale describes the size of the regions where both species are present and, therefore, natural selection can act. These regions occur at the boundaries between the domains, so we refer to this length scale as the boundary width,~$L_{\rm{b}}$. Domain boundaries have finite size because the constant exchange of migrants across the boundary is balanced by their extinction due to genetic drift; $L_{\rm{b}}=mN$ from Ref.~\cite{hallatschek2009}. The intuition behind these results and the relationships among the length scales are discussed in the Supplemental Material~\cite{supplement}.

We will now assume that the population is sufficiently close to the demixing phase transition so that the distance between domain boundaries~$L_{\rm{d}}$ is much greater than~$L_{\rm{n}}$, as it is commonly observed experimentally~\cite{muller2014, momeni2013, momeni2013spatial}. When~$L_{\rm{d}}\gg L_{\rm{n}}\gg L_{\rm{b}}$, one can solve Eq.~(\ref{eq:fmodel}) near the domain boundary located at~$x=0$ by assuming that~$f_{\rm{A}}(x)$ is a step function, i.e.~$f_{\rm{A}}(x)=1$ for~$x<0$ and~$f_{\rm{A}}(x)=0$ for~$x>0$. The solution yields

\begin{equation}
	n_{\rm{B}}(x)-n_{\rm{A}}(x)=\mbox{sgn}(x)\frac{p}{d}(1-e^{-|x|/L_{\rm{n}}})\approx xp/(dL_{\rm{n}})
	\label{eq:pg_boundary}
\end{equation}

\noindent for~$|x|\ll L_{\rm{n}}$, where~$\mbox{sgn}(x)$ is the sign function. 

Since the selection term in Eq.~(\ref{eq:fmodel}) vanishes when~$f_{\rm{A}}(1-f_{\rm{A}})=0$, only the fitness differences at the domain boundary~(when~$|x|\approx L_{\rm{b}}$) affect population dynamics. Near the boundary~$f_{\rm{A}}(x)\approx1/2-x/L_{\rm{b}}$. Hence, by eliminating~$x$ from Eq.~(\ref{eq:pg_boundary}), we can again recast the selection term in the form~$s_{\rm{eff}}f_{\rm{A}}(1-f_{\rm{A}})(1/2-f_{\rm{A}})$, where the effective strength of selection~$s_{\rm{eff}}\sim (p/d)(L_{\rm{b}}/L_{\rm{n}})$ is reduced by a factor of~$L_{\rm{b}}/L_{\rm{n}}=mN/\sqrt{D/d}$ compared to the model with~$D=0$.

%figure 3 can become figure 2b
\begin{figure}[h]
\includegraphics[width=\columnwidth]{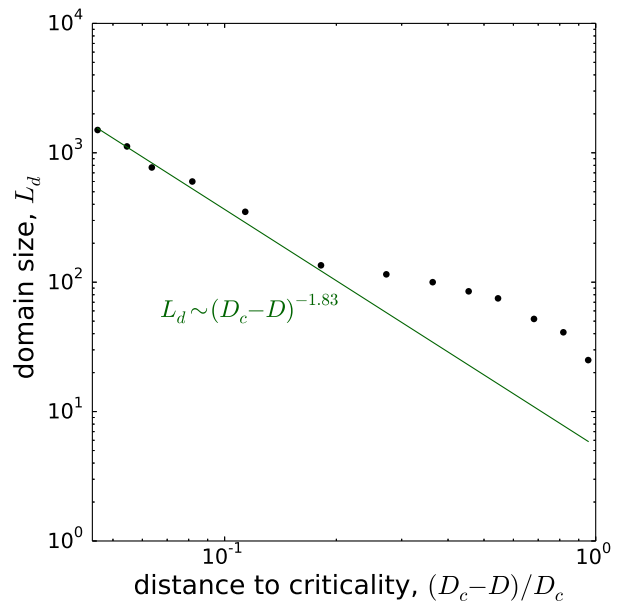}
\caption{The scale of species intermixing is controlled by the underlying nonequilibrium phase transition. The dots are the simulation data, and the line is the fit of the expectation that~$L_{\rm{d}} \sim (D_{c} - D)^{-v_{\perp}}$. The theory and simulations agree close to the phase transition when domains are large. For small~$D$,~the size of the domains is determined by the dynamics studied in Ref.~\cite{korolev2011}. \label{fig3}} 
\end{figure}

Our finding that higher diffusivities of the public goods reduce the effective strength of selection explains the decrease of~$H$ with~$D$ in Fig.~\ref{fig1} and provides a way to estimate the critical diffusivity~$D_{\rm{c}}$ above which mutualism is lost. Indeed, when~$D=D_{\rm{c}}$, we expect that the strength of mutualism~$S$ approaches its critical value as well. Thus,~$S_{\rm{c}}=s_{\rm{eff}}mN^2\sim pm^2N^3/\sqrt{D_{\rm{c}}d}$, and~$D_{c} \sim s^2m^4N^6d^{3}$, where~$s=2p/d$ is the strength of selection in the model without public good diffusion. Surprisingly, we find that population density and migration have a much stronger effect on the critical nutrient diffusivity than natural selection. Our simulation results are in excellent agreement with these predictions (Fig.~\ref{fig2}).% Reduced selection due to public good diffusion might also explain the residual species demixing observed even for obligate mutualists~\cite{muller2014}.

Since the model with~$D>0$ is equivalent to that with~$D=0$ provided the strength of selection~$s$ is renormalized, many of the results from the evolutionary game theory can be generalized for microbial communities with diffusible public goods. The size of the domains formed by the species~$L_{\rm{d}}$ is of particular interest because it is used in the experiments to quantify the degree to which the two species benefit from their mutualistic interactions. Previous studies suggested that~$L_{\rm{d}}\sim D^{1/4}$~\cite{muller2014} or~$L_{\rm{d}}\sim D^{1/5}$~\cite{momeni2013}; however, we find that such scalings are unlikely because~$L_{\rm{d}}$ becomes large only close to the underlying phase transition, where~$L_{\rm{d}}\sim(D_{\rm{c}}-D)^{-\nu_{\perp}}$ for~$D<D_{\rm{c}}$; see Fig.~\ref{fig3}. The exponent~$\nu_{\perp}$ is that of a correlation length and is determined by the universality class of the phase transition. In our model, any species asymmetry results in DP universality class and the extinction of one of the two species when mutualism is lost. If all model parameters are the same for the two species, the dynamics is in DP2 universality class and the loss of mutualism results in spatial demixing~\cite{hinrichsen2000}. The nature of this phase transition does not depend on whether one varies the diffusivities of the public goods, their external concentrations~\cite{muller2014}, or any other model parameters.

\begin{figure}[h]
\includegraphics[width=\columnwidth]{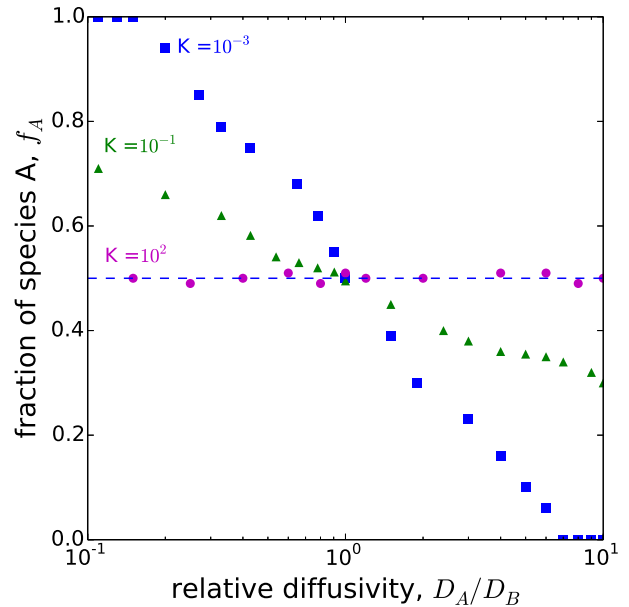}
\caption{For large~$K$, the model is linear and the differences in the nutrient diffusivities have no affect on the relative species abundances~(magenta dots). For lower~$K$, fitness nonlinearity increases, and the species with the lower diffusivity dominates~(green triangles). This trend continues as~$K$ is decreased further, but, in addition, species coexistence and mutualism are lost when public goods diffusivities become too unequal~(blue squares). \label{fig4}} 
\end{figure}

Finally, we investigated whether species can take advantage of each other by changing the diffusion constant of their public goods. Upon repeating the steps leading to Eq.~(\ref{eq:fmodel}) but for~$D_{\rm{A}}\ne D_{\rm{B}}$ one still finds that the resulting equation is invariant under the exchange of species labels, i.e.~$f_{\rm{A}}\to1-f_{\rm{A}}$. Hence, if only the public good diffusivities are different between the species, then none of the species is expected to dominate the other. This conclusion however holds only when the fitnesses are linear functions of the nutrient concentrations, i.e.~$K=\infty$. For lower values of~$K$, we find that the species producing public goods that diffuse more slowly dominates the other species (Fig.~\ref{fig4}). As~$K$ is decreased further, the population undergoes the demixing phase transition described above, and one of the species becomes extinct.

The effects of fitness nonlinearities and public good diffusivities can be easily understood by considering the population dynamics close to the domain boundary. The dominant species is determined by whether species A is more likely to invade the space occupied by species B or species B is more likely to invade the space occupied by species A. To make the argument more clear let us assume that~$D_{\rm{A}}=0$ and~$D_{\rm{B}}=\infty$, then the concentration of public good B is the same everywhere while the concentration of public good A is high inside the domain comprised of species A and zero outside. As a result, the fitness of species A is the same everywhere, while the fitness of species B is low in its own domain and high in the domain occupied by species A. The nonlinearity in Eq.~(\ref{eq:fitness}) makes fitness changes at low nutrient concentrations much more pronounced than at high nutrient concentrations. Thus, the advantage that the species B has over A in the domain occupied by species A (where~$n_{\rm{A}}$ is high) is smaller than the advantage that species A has over B in the domain occupied by species B~(where~$n_{\rm{A}}=0$). As a result, species A with lower public good diffusivity dominates species B in agreement with the simulations~(Fig.~\ref{fig4}). 

In summary, we demonstrated that the main effect of public good diffusion is the reduction of the effective strength of natural selection, which can lead to the loss of mutualism via a nonequilibrium phase transition. The distance to this phase transition controls the size of the domains formed by the species, a quantity of prime interest in empirical studies. In addition, differences in the diffusivities of the public goods could have a profound effect on the population dynamics. The effect of these differences depends on the fitness and other nonlinearities and results in the selective advantage for one of the species. Our work together with Ref.~\cite{korolev2011} provides a theory for the phenomena observed in recent experimental studies~\cite{muller2014, momeni2013, momeni2013spatial} and could potentially explain why cooperatively growing microbes modulate the diffusivities of their public goods~\cite{kummerli2014}.

\begin{acknowledgments}
This work was supported by the startup fund from Boston University to KK. Simulations were carried out on Shared Computing Cluster at BU.
\end{acknowledgments}

\setcounter{equation}{0}
\setcounter{figure}{0}
\renewcommand{\thefigure}{S\arabic{figure}}
\renewcommand{\theequation}{S\arabic{equation}} 

\section*{Appendix : Supplemental material \\ Public good diffusion limits microbial mutualism }

In this supplemental material, we clarify and expand some of the statements made in the main text. Specifically, we include a detailed discussion of the length scales in population dynamics and nutrient diffusion, illustrations of our arguments for the renormalization of fitness and the effects of fitness nonlinearities, description of the transition to mutualism on a global scale, discussion of alternative models of public good production and consumption, comparison of mutualism to balancing frequency-dependent selection, and the details of our computer simulations.\\

\textbf{Four length scales: $\bm{L}$, $\bm{L_{\rm{d}}}$, $\bm{L_{\rm{b}}}$, and~$\bm{L_{\rm{n}}}$}\\

\begin{figure}[h]
\begin{center}
\includegraphics[width=\columnwidth, height = 4in]{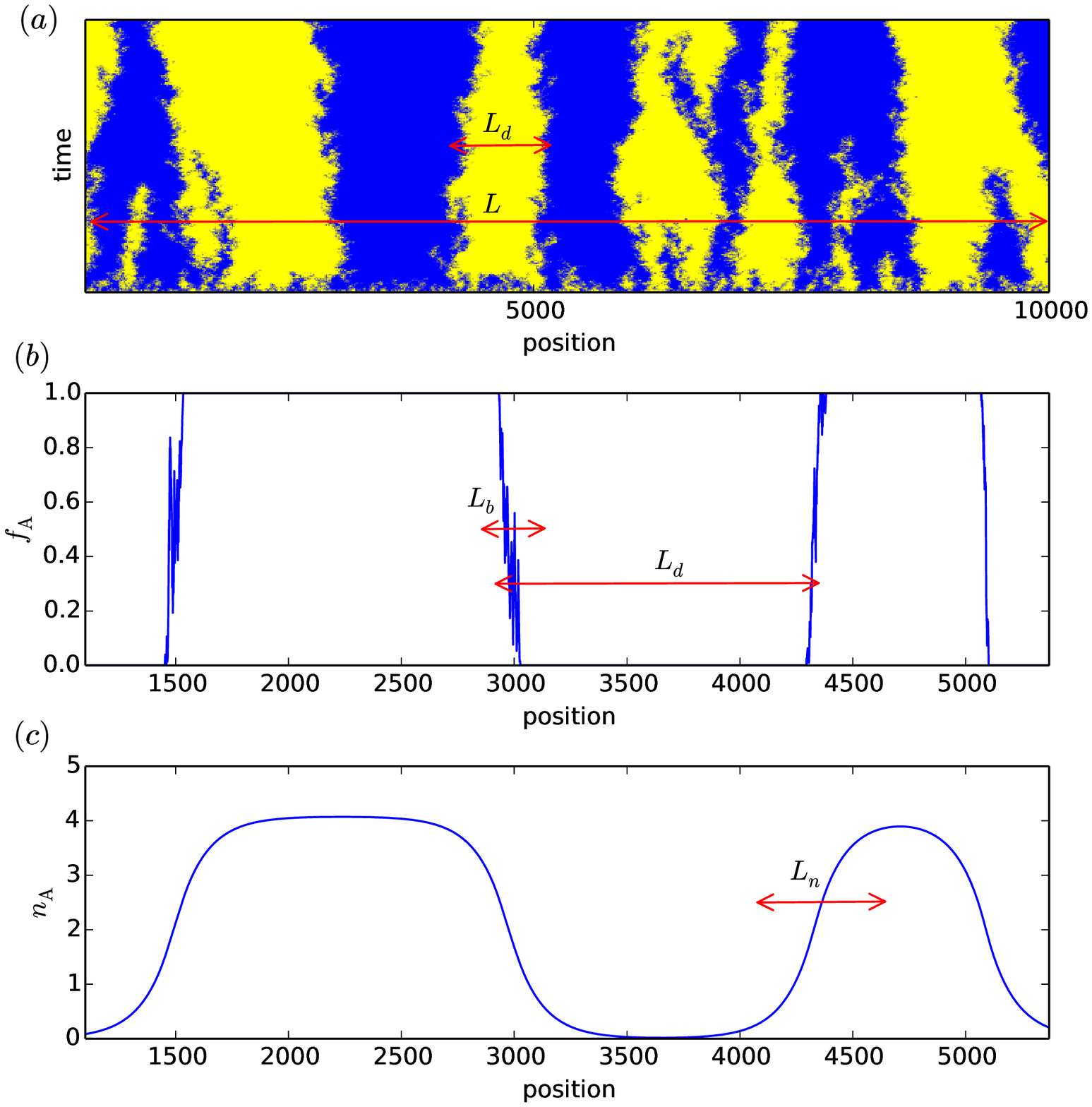}
\end{center}
\caption{This figure illustrates four main length scales under consideration. (a)~The top panel shows how the spatial distribution of species evolves in time on the scale of the whole population~$L$. The species are shown with different colors. For large~$D$, mutualism is weak and species demix into domains dominated by one of the two species. The size of these domains is $L_{\rm{d}}$. (b)~The next panel zooms in on the scale of a single domain and compares the domain size $L_{\rm{d}}$ and the size of domain boundaries $L_{\rm{b}}$. The latter length characterizes the region between the two domains where both species are present simultaneously. (c)~The final panel shows the distribution of the public goods produced by the species shown in (b). For large~$D$, public goods diffuse well outside the domain boundaries before being consumed or degraded. The length scale over which public goods concentrations change is $L_{\rm{n}}$. Here, $p =0.0005$, $d = 1$, $m=0.1$, $N=500$, and~$D = 150000$.}
\label{cartoon} 
\end{figure}

In two-way cross feeding, species fractions and nutrient concentrations change on four key length scales that we describe below. These are the system size~$L$, the length on which species fractions remain constant~$L_{\rm{d}}$, the length scales over which species fractions change~$L_{\rm{b}}$, and the length scale on which the concentration of public goods change~$L_{\rm{n}}$.\\

The first length scale is the system size~$L$, which is the number of islands in our simulations or the circumference of a microbial colony in the experiments. The manuscript assumes that this length scale is much larger than any other length scale we consider. When that is not the case, population dynamics can change either because species can easily migrate across the whole population making it effectively well-mixed rather than spatially structured or nutrients can easily diffuse across the whole population making nutrient concentrations homogeneous rather than constrained by the diffusive transport. The latter situation is carefully discussed later in this Supplemental Material.\\    

The second length scale is the size of the spatial regions occupied exclusively by one of the species. We term these regions domains and denote the corresponding length scale as the size of the domains~$L_{\rm{d}}$. In Fig.~\ref{cartoon}a, domains are represented by patches of the same color and, in Fig.~\ref{cartoon}b, as regions where~$f_{\rm{A}}$ equals either~$0$ or~$1$. Note that the different domains have different sizes, i.e. the size of the domains is stochastic. Therefore,~$L_{\rm{d}}$ represents the average size of the domains. Also, we define~$L_{\rm{d}}$ only when the domain size distribution has reached equilibrium. When no equilibrium exists, e.g., when complete demixing occurs,~$L_{\rm{d}}$ is time dependent, and its equilibrium value cannot be defined.\\

The third length scale is the boundary width $L_{\rm{b}}$, which describes the size of the regions that separate the domains. This length scale is distinct from~$L_{\rm{d}}$ because domains are much larger than the transitions between them. Moreover, the boundaries between the species are not infinitely sharp, but have a certain width~(Fig.~\ref{cartoon}). Finite boundary width does not require any selection or mutualism and is observed even in neutral models. The two processes that determine the boundary width $L_{\rm{b}}$ are migration and genetic drift. Migration makes the boundaries wider since more migrant exchange between the domains (when migration is zero the boundaries are infinitely sharp). Genetic drift is also important. Without drift, population dynamics are purely diffusive, which leads to a spatially homogeneous state without domains. Very loosely speaking, the survival time of an individual of species B in the domain of species A is~$N$ because of genetic drift and the maximal distance that that individual can travel in that time is~$m$~times~$N$, i.e.~$L_{\rm{b}} = mN$. This argument is not rigorous, and the actual dynamics is much more intricate as explained in Ref.~\cite{hallatschek2009}.\\

The fourth length scale~$L_{\rm{n}}$ is the distance that public goods diffuse before being absorbed. Since the life time of public goods is~$1/d$, they typically spread over a distance~$L_{\rm{n}}\sim\sqrt{D/d}$. This nutrient length scale can also be immediately obtained from Eq.~(4) in the main text, where it appears in the exponent. Another interpretation of~$L_{\rm{n}}$ is the distance over which changes in the species fractions cause changes in the concentrations of the public goods. For small diffusivities, nutrients are absorbed where they are produced, and nutrient concentrations closely follow species fractions. For large diffusivities, the concentrations of public goods change more gradually. As a result, public goods produced by species A diffuse well into the domain occupied by species B (Fig.~\ref{cartoon}c). \\

\textbf{Expansion of Eq.~(5) to the first order}\\

One way to see that nutrient diffusion inhibits mutualism in to consider the expansion of Eq.~(5) for small~$D$. The first order of the expansion contributes a term~$-2pDd^{-2}f_{\rm{A}}(1-f_{\rm{A}})\partial^2 f_{\rm{A}}/\partial x^2$ in addition to the term that we obtained by setting~$D=0$. Biologically, this new term implies that species~A is at a disadvantage in the locations where its density has a local minima because species~B receives extra public goods from the nearby regions with larger concentration of species~A. Mathematically, this term has a similar form to the second-derivative term describing species migration in~Eq.~(1), but contributes with the opposite sign. In the language of field theory, public good diffusion renormalizes species migration to a lower value. Since the strength of mutualism~$S$ increases with the migration rate~$m$, we expect that public good diffusion should favor species demixing.\\

\textbf{Public goods diffusion renormalizes the selection for coexistence} \\

\begin{figure}
\begin{center}
\includegraphics[width=\columnwidth]{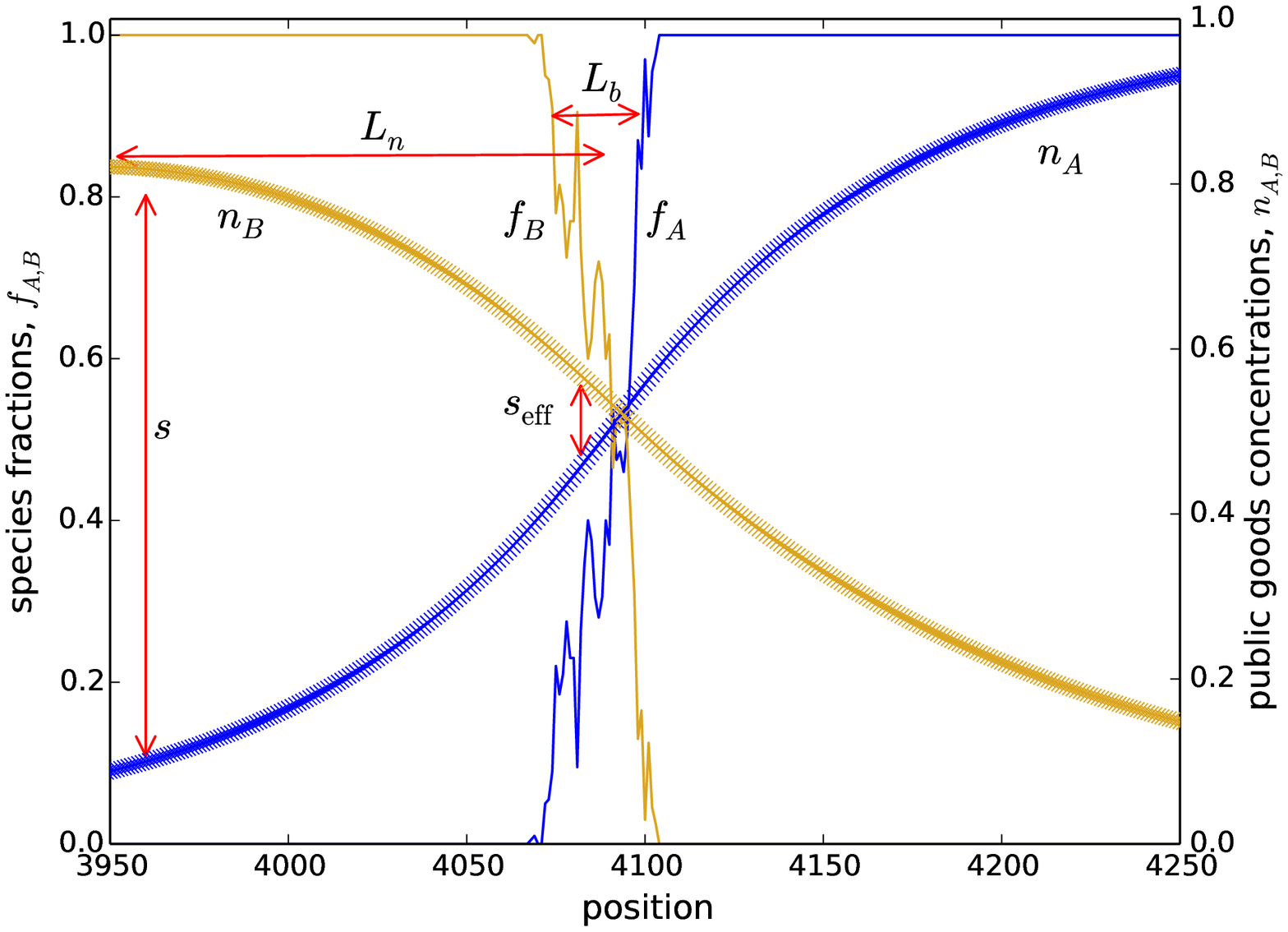}
\end{center}
\caption{This figure illustrates our argument that~$s_{\rm{eff}}\sim s\frac{L_{\rm{b}}}{L_{\rm{n}}}$. Selection occurs only in the region of size~$L_{\rm{b}}$, where the differences in nutrient concentrations are small. This leads to a reduced difference in species fitness~$s_{\rm{eff}}$ compared to the maximally possible fitness difference~$s$. \label{fig:renormalization}} 
\end{figure}

\begin{figure}
\begin{center}
\includegraphics[width=\columnwidth]{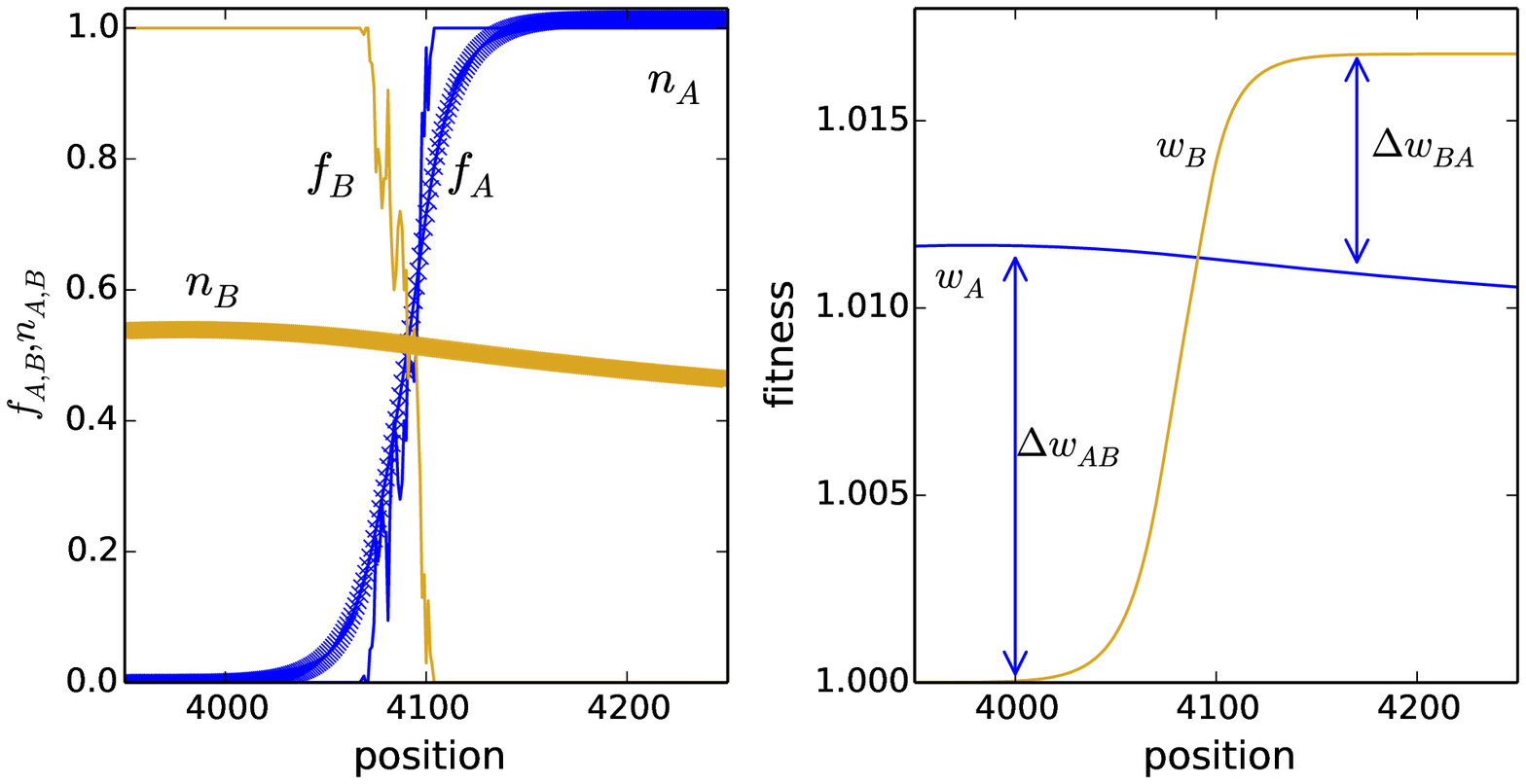}
\end{center}
\caption{The left panel shows the spatial distribution of public goods close to the domain boundary, when the diffusivities of the public goods are different. The right panel shows the effect of nutrient distributions on the fitness of the species, when fitness saturates at high nutrient concentrations. \label{fig:nonlinear}} 
\end{figure}

Our arguments for the dynamics near the boundary between the two species are illustrated in Fig.~\ref{fig:renormalization} for the reduction in the selection for coexistence due to public good diffusion and in Fig.~\ref{fig:nonlinear} for the effects of fitness nonlinearities. Note that reduced selection due to public good diffusion might also explain the residual species demixing observed even for obligate mutualists; see figure 2C in Ref.~[3] of the main text.\\

\textbf{Mean fitness is an alternative order parameter} \\

\begin{figure}
\begin{center}
\includegraphics[width=\columnwidth]{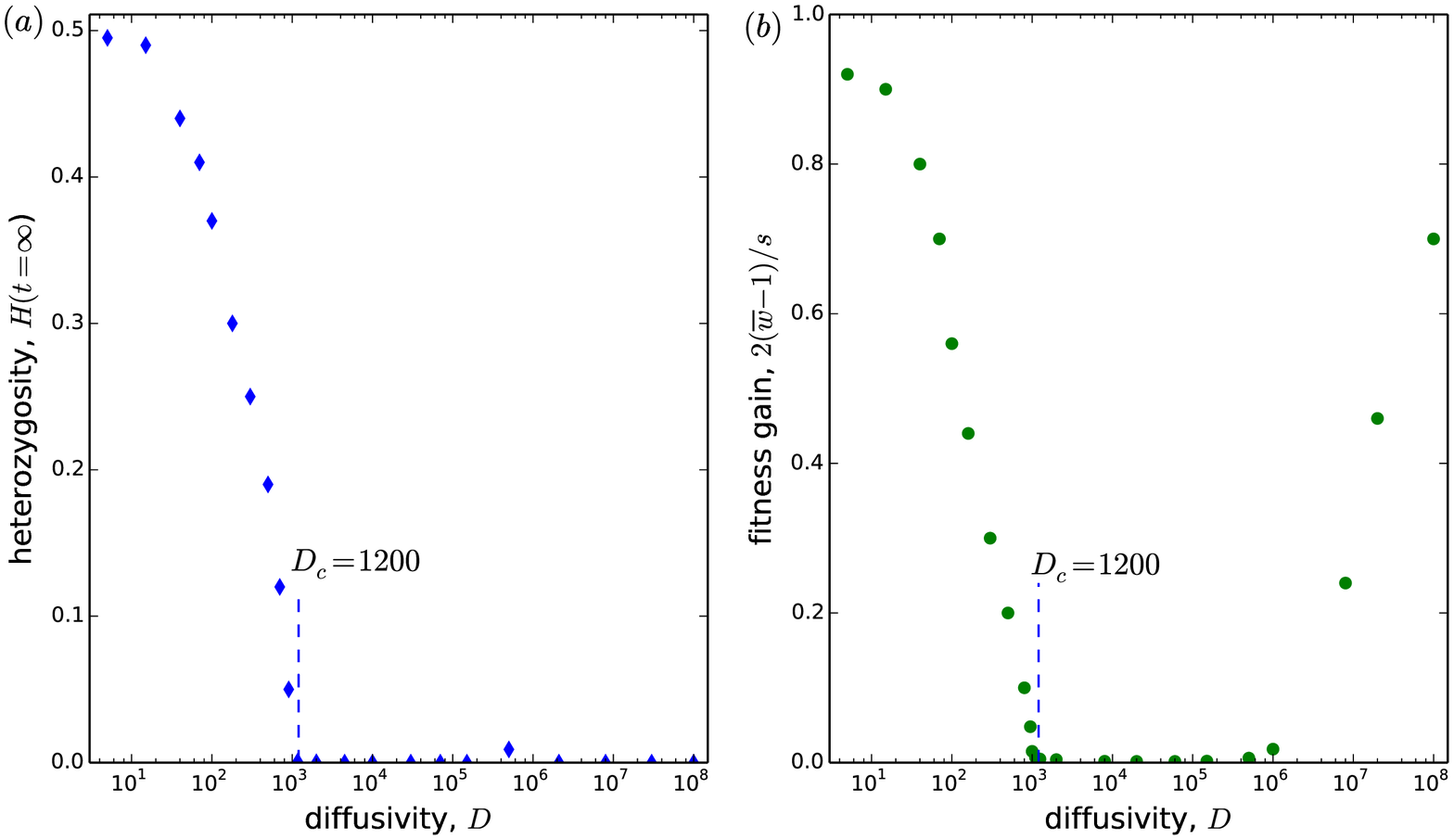}
\end{center}
\caption{Both local heterozygosity and fitness gain due to mutualism become zero at the critical diffusivity. (a)~Local heterozygosity~$H$ decreases with nutrient diffusivity~$D$. The data is the same as in Fig.~1 of the main text. As we discuss in the main text, $H > 0$ for $D < D_{\rm{c}}$, and $ H = 0$ for $D \geq D_{\rm{c}}$. (b)~The fitness gain due to mutualism $4( \bar{w} - 1)/s$ is shown. Here, $\bar{w}$ is the arithmetic mean of the fitnesses of all organisms in the population. The fitness gain equals~$1$ when species take maximally possible advantage of the mutualism and~$0$ when mutualism is lost. Both local heterozygosity and fitness gain show similar trends for nutrient diffusivities that are not too large. In particular, mutualism is lost above the critical diffusivity. However, when~$D$ becomes so large that the nutrient length scale~$L_{\rm{n}}$ becomes comparable to the system size~$L$, mutualism recovers even though~$H$ remains zero. This mutualism occurs on a global rather than on a local scale. Here, the simulation parameters are the same as in Fig.~1 of the main text.\label{fig:mean_fitness}} 
\end{figure}

Local heterozygosity~$H$ has been previously used as an order parameter describing the mutualistic phase~\cite{korolev2011}. The connection between~$H$ and mutualism is easy to understand in the standard game theory because only species present in the same location can interact. Indeed, there are no species interactions in a demixed state, and, therefore, populations with~$H=0$ cannot support mutualism. This connection between~$H$ and mutualism is less clear when public goods can diffuse because, even in the demixed state, species can benefit from each other if they can exchange public goods through diffusion. Since both the domain size~$L_{\rm{d}}$ and the nutrient exchange length scale~$L_{\rm{b}}$ increase with~$D$, it is important to demonstrate that the loss of local heterozygosity indeed implies the loss of mutualism.\\

The degree of mutualism is directly linked to the mean fitness of the population~$\bar{w}$. When mutualism is successful, the mean fitness will be larger than one~(about~$1+s/4$), but, when the mutualism fails due to demixing,~$\bar{w}$ will be close to one. Unlike~$H$, which quantifies species coexistence, mean fitness quantifies the affect of mutualism on cell growth. Thus~$\bar{w}$ is a more robust measure of mutualism than~$H$ because it is directly applicable for all values of~$D$. For convenience, we normalize~$\bar{w}$ by its maximal possible value and define the fitness gain due to mutualism as $4(\bar{w} - 1)/s$, which varies between 0 and 1.\\ 

Figure~\ref{fig:mean_fitness} shows how both local heterozygosity and the fitness gain due to mutualism change with nutrient diffusivity~$D$ for the same parameters as Fig.~1 in the main text. For~$D$ that is not too large, both measures of mutualism show similar trends. In particular, one can clearly see that both~$H$ and the fitness gain vanish at~$D_{\rm{c}}$. Thus, the phase transition described in the main text is the loss of both mutualism and coexistence. The behavior for very large~$D$ is discussed in the next section.\\

\textbf{Transition to global mutualism}\\

Throughout the main text we assume that the system size (e.g. the number of islands in our simulations) is much greater than the typical distance that the nutrients diffuse before being consumed. This separation of length scales is indeed observed in microbial experiments. Typically, the circumference of a microbial colony is between~5mm and 10cm, while the nutrient diffusion length was estimated to be between 30$\mu$ and~700$\mu$~\cite{muller2014, momeni2013}. The large system size limit is also necessary to properly study phase transitions in our simulations.\\ 

The behavior in the other limit, when~$L_{\rm{n}}\gg L$ is also interesting. In this limit, nutrient concentrations are spatially homogeneous and nutrient exchange occurs on a global rather than on a local scale. Although such fast nutrient transport cannot typically occur through diffusion, fluid flow driven by convection, mixing, or turbulence can in principle create a regime of global nutrient exchange. Dynamics in this regime are very different from what we describe in the main text because the strength of frequency-dependent selection depends on the mean fractions of the species in the entire population. Thus, there is a strong selection for species coexistence on a global scale even though they demix locally.\\

Simulation results shown in Fig.~\ref{fig:mean_fitness} support this intuition and illustrate the transition from local to global mutualism. For~$D$ that is not too large,~$L_{\rm{n}}\ll L$ is satisfied. Therefore, both~$H$ and the fitness gain due to mutualism decrease with~$D$ until they vanish at~$D_{\rm{c}}$. The local heterozygosity remains zero as~$D$ increases by orders of magnitude~(beyond what is physically possible). The fitness gain, however, remains zero only while $L_{\rm{n}}\ll L$. When the diffusion is fast enough to make nutrient concentrations homogeneous across the population, the regime of global mutualism starts, and the fitness gain becomes nonzero.\\

In the regime of global mutualism one finds that species demix locally until there are just two domains left: one of species A and the other of species B. The relative size of these domains fluctuates around~$f^*$, i.e. the mutualism is preserved even though, the coexistence is lost locally. Because nutrients are shared globally, population dynamics in this regime are similar to those of a well-mixed population of size~$NL$. The only difference is that the competition between the species is restricted to the boundary between the two domains. This spatial localization of competition results in quantitative but not qualitative differences in the population dynamics between spatial and well-mixed populations.\\

When~$L\gg L_{\rm{n}}$ and~$D>D_{\rm{c}}$, local demixing will also eventually leave only two domains: one of species A and one of species B just like in the regime of global mutualism. Unlike that regime however, the frequency-dependent selection will be strongly suppressed. Indeed, the concentration of nutrients at the boundary where the competition occurs will be the same regardless of the global fraction of species A and B as long as the domains occupied by the species are larger than~$L_{\rm{n}}$. Only when the domain of one of the species shrinks below~$L_{\rm{n}}$, the balancing frequency-dependent selection becomes nonzero due to the lack of the public goods produced by the shrinking species. To put this another way, there is no pressure for the boundary between the species to separate them in equally sized domains because the species at the boundaries get their public goods from a region of size~$L_{\rm{n}}$, not~$L$, and, within the distance of~$L_{\rm{n}}$ from the boundary, the species are about equally abundant. As a result, mutualism in this regime is not global and is not stable. \\

\begin{figure}
\begin{center}
\includegraphics[width=\columnwidth]{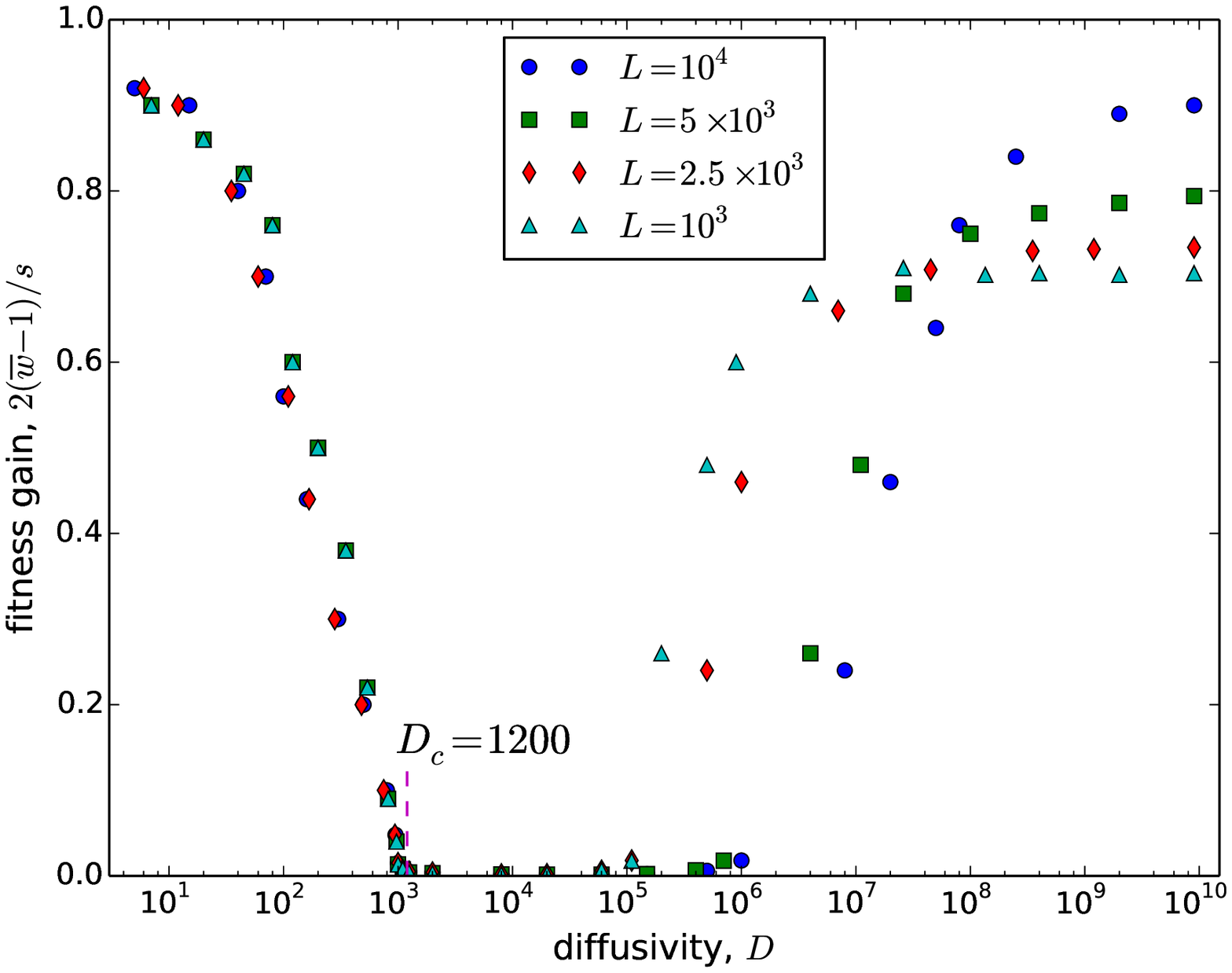}
\end{center}
\caption{Transition to global mutualism depends on the system size. Similar, to Fig.~\ref{fig:mean_fitness}, the loss of mutualism due to local demixing occurs at~$D_{\rm{c}}$ for all four system sizes. The fitness gain remains zero as diffusivity increases until~$L_{\rm{n}}=\sqrt{D/d}$ becomes comparable to~$L$, and global mutualism sets in. As expected, the diffusivity, at which the transition to global mutualism occurs, increases with the system size. Note that the value of fitness gain at~$D=\infty$ approaches~$1$ as~$L$ increases because genetic drift is suppressed in large populations. Here, $p=0.0005$, $d=1$, $N=200$, and~$m=0.1$. \label{system_size}} 
\end{figure}

The effect of the system size~$L$ on the transition to global mutualism is illustrated in Fig.~\ref{system_size}. For all system sizes, we observe the loss of mutualism at~$D=D_{\rm{c}}$ followed by a region of no mutualism at higher diffusivities. However, when~$L_{\rm{n}}=\sqrt{D/d}$ becomes comparable to the system size, there is a transition to global mutualism. As we expect, this second transition occurs at larger~$D$ for larger system sizes. In consequence, the region of no mutualism increases with the system size, and becomes infinitely large as~$L\to\infty$ indicating that the loss of mutualism at~$D_{\rm{c}}$ is a true phase transition.\\  

Note that the transition from local to global mutualism that we discussed here is quite different from the often observed effect that spatial structure promotes coexistence. For example, Ref.~\cite{kerr2006} found that bacteria playing an equivalent of Rock-Paper-Scissors game cannot coexist in a well-mixed population, but maintained global coexistence in a spatially structured population even though local diversity was lost. Although fast nutrient transport also leads to global coexistence at the expense of local diversity, the mechanism of coexistence is very different. Indeed, global nutrient exchange makes a spatially structured population of two-way cross-feeders behave essentially the same as a well-mixed population.\\

\textbf{Alternative models of public goods dynamics} \\

In the main text, we used the following model of public goods production, consumption, and decay

\begin{equation}
\frac{\partial n_{{A}}}{\partial t} = D_{{A}}\frac{\partial^{2}n_{{A}}}{\partial x^{2}} + p_{{A}} f_{\rm{A}} - d_{{A}} n_{{A}}.
\label{eq:pg} 
\end{equation}

\noindent This is perhaps the simplest model that assumes that public good~$A$ is constantly produced by species~A at a constant rate and either both species consume the public good at the same rate or nutrient decay is primarily caused by external factors  other than the consumption by the species; for example, due to diffusion outside the layer of actively growing cells or degradation. Deviations from these assumptions will make~$p_{{A}}$ and~$d_{{A}}$ depend on~$n_{{A}}$,~$n_{{B}}$, and~$f_{\rm{B}}$. Additional dependence on~$f_{\rm{A}}$ is not necessary since~$f_{\rm{A}}=1-f_{\rm{B}}$. Next, we consider three different models that incorporate such dependencies and show that our results are robust to these changes.\\

\underline{Model~1.} In this model, public good production saturates when the producing species is very abundant. Biologically, Model~1 reflects cellular regulation that allows the production of public goods only when they are needed.  

\begin{equation}
\frac{\partial n_{{A}}}{\partial t} = D_{{A}}\frac{\partial^{2}n_{{A}}}{\partial x^{2}} + p_{{A}} \frac{f_{\rm{A}}}{(1+f_{\rm{A}})} - d_{{A}} n_{{A}}
\end{equation}

\noindent Note that this model is nonlinear in species fractions.\\

\underline{Model~2.} In this model, public goods are consumed only by species~B, and there is no other mechanism of public goods decay.
 
\begin{equation}
\frac{\partial n_{{A}}}{\partial t} = D_{{A}}\frac{\partial^{2}n_{{A}}}{\partial x^{2}} + p_{{A}} f_{\rm{A}} - d_{{A}} n_{{A}}f_{\rm{B}}
\end{equation}

\underline{Model~3.} In this final model, nutrient is preferentially consumed by species~B and can also decays due to external factors. 
\begin{equation}
\frac{\partial n_{\rm{A}}}{\partial t} = D_{\rm{A}}\frac{\partial^{2}n_{\rm{A}}}{\partial x^{2}} + p_{\rm{A}} f_{\rm{A}} - d_{\rm{A}} n_{\rm{A}} - f_{\rm{B}} n_{\rm{A}}
\end{equation}

\noindent Note that the last term in both Models~2 and~3 introduces a nonlinearity because it is a product of species fractions and public good concentrations.\\
 
\begin{figure}
\begin{center}
\includegraphics[width=\columnwidth, height = 1.5in]{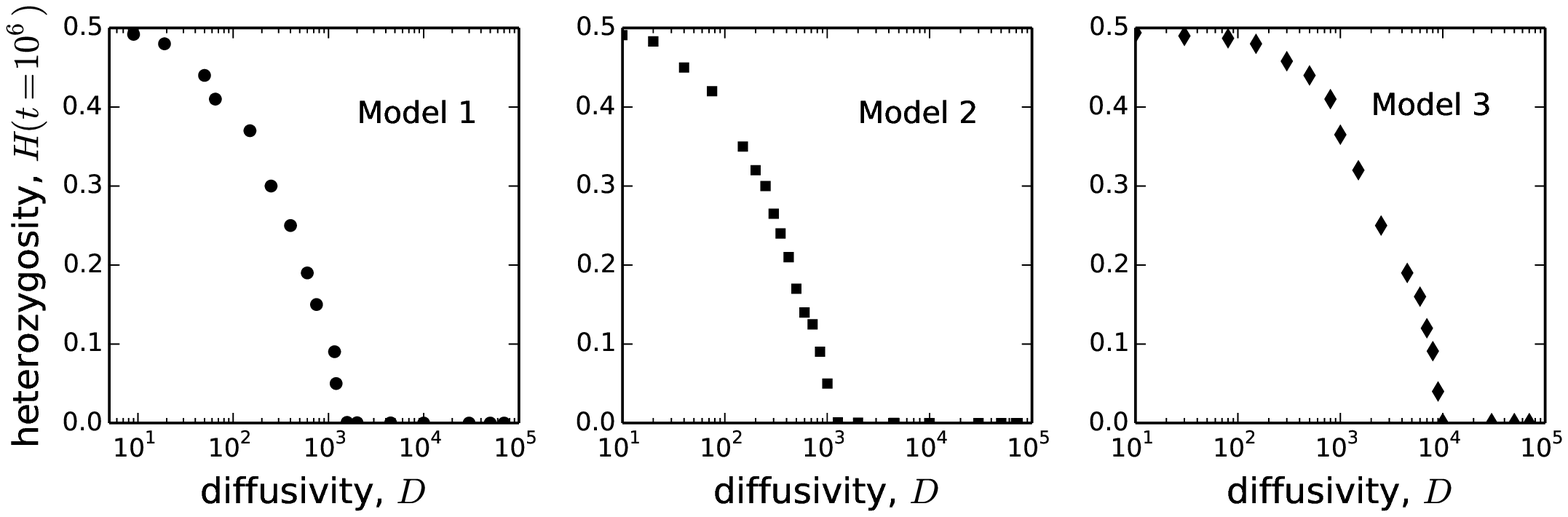}
\end{center}
\caption{Loss of coexistence with increasing public goods diffusivity for the three models discussed in the text. The behavior of the heterozygosity in all three models is consistent with the simpler model analyzed in the main text. The parameters used in the simulations are as follows. Model~1: $N=200$, $m=0.1$, $p=0.0005$, and~$d=1$. Model~2: $N=200$, $m=0.1$, $p=0.0005$, and~$d=1$. Model~3: $N=500$, $m=0.003$, $p=0.5$, and~$d=10$. \label{models}} 
\end{figure}

In simulations, we observe the loss of coexistence with increasing public goods diffusivity for all of these models; see Fig.~\ref{models}. Except for the fact that nonlinearities affect competition between species with different public goods diffusivities, we did not find any qualitative difference in the behavior of these models and the model studied in the main text.\\

\textbf{Effects of asymmetries in public goods dynamics}\\

We now turn to the effects of species asymmetries on the population dynamics. When~$D=0$ and the model reduces to that of local frequency-dependent selection, the asymmetries in nutrient production and decay rates result in the selection term of the form~$sf_{\rm{A}}(1-f_{\rm{A}})(f^{*}-f_{\rm{A}})$, where the preferred fraction~$f^{*}$ no longer equals~$1/2$. Population dynamics in this limit have been previously analyzed in Ref.~\cite{korolev2011} with the main conclusion that species asymmetry substantially weakens mutualism and species~A is favored if~$f^{*}>1/2$ while species B is favored otherwise. We find that~$D>0$ does not alter these results.\\ 

\textbf{Frequency-dependent selection in population genetics}\\

In the main text, we focused on mutualism as the force maintaining coexistence between different species. Coexistence among genotypes has been extensively studied in population genetics, and, in this section, we draw some parallels to that field. Most of the population genetics literature focuses on the maintenance of polymorphism due to site-specific or local adaptation. In the context of our model, this would correspond to different islands always favoring either species A or species B. This situation however is quite different from mutualism, and we do not think that a simple analogy can be made between these two alternative mechanisms to maintain polymorphisms, i.e. local adaptation vs. local frequency-dependent selection. Indeed, local adaptation fails at high migration rates, while the stability of coexistence always increases with migration for local frequency-dependent selection.\\

Without public good diffusion, mutualism between two species is however analogous to frequency-dependent selection in population genetics because these two phenomena are described by the same mathematical model. An example of frequency-dependent selection especially relevant to our discussion of mutualism is the $t$-allele in house mice. This allele is under balancing selection, but is observed at frequencies much smaller than predicted by a deterministic model. Lewontin~\textit{et al.} first proposed that the reduced prevalence could be due to stochastic effects~\cite{lewontin1960}. This idea was further refined to account for migration in Ref.~\cite{levin1969}, and a very definitive treatment was presented by Durand~\textit{et al.}~\cite{durand1997}. This last paper found that there is a critical migration below which the $t$-haplotype becomes extinct (for much higher migration the deterministic approximation works well). This result parallels that of Ref.~\cite{korolev2011} because~$S_{\rm{c}}\sim sm^2N^4$ and the phase transition leading to the loss of coexistence can occur by reducing selection, population size, or migration. The analysis by Durand~\textit{et al.} was however done for an island model, not a stepping-stone model. \\

\textbf{Details of simulations} \\

Simulations were implemented based on the Wright-Fisher model of population genetics. Concretely, we first computed the expected species fractions in the next generation and then obtained the actual number of species in each deme via binomial sampling with~$N$ trials. We considered spatial structure of a quasi-one-dimensional population because the growth of cells in a microbial colony is confined to a very narrow region~(a few cell widths) at the colony edge~\cite{lavrentovich_2013}.\\
 
In Fig.~3 of the main text, we obtain~$L_{\rm{d}}$ as follows. First, we compute~$H(t,x)$, a two point correlation function that estimates the probability of sampling two different species a distance $x$ apart:
  
\begin{equation}
 H(t,x) = \langle f(t,0)[1-f(t,x)] + f(t,x)[1-f(t,0)] \rangle.
\end{equation}

When population reaches a steady state,~$H(t,x)$ becomes independent of time, and we denote this equilibrium heterozygosity as~$H(x)$. The domain length is then obtained by fitting the exponential decay of~$H(x)$ at large~$x$, i.e.~$H(x) - H(\infty) \sim e^{-{x}/{L_d}}$, which is expected for all~$D$ other than~$D=D_{\rm{c}}$~\cite{korolev2011, hinrichsen2000}.\\
 
Data in all figures is averaged over~$20$-$25$ replicates. In Figs.~1, 3, and~4 of the main text, $N = 200$, $m = 0.1$, $p = 0.001$, $d = 1$, and $K = 1$. In Fig.~2 of the main text,~$K=1$ while other parameters are given in the figure panels. In Figs.~1, 2, and~3 of the main text, the habitat consisted of~$10^{4}$ islands and was observed after~$10^{6}$ generations starting from a well-mixed state. In Fig.~4 of the main text, the habitat consisted of~$500$ islands and was observed after~$10^{5}$ generations starting from a well-mixed state.

\end{document}